\documentclass[a4paper,fleqn,usenatbib]{mnras}

\usepackage[utf8]{inputenc}
\usepackage{times}
\usepackage{graphicx,bm,amssymb}
\usepackage{epsfig}
\usepackage{natbib}
\usepackage{psfrag}
\usepackage{graphicx}
\usepackage{ulem}
\usepackage{tikz}
\usepackage{etaremune}
\usepackage{pifont}
\usepackage{colortbl}
\usepackage{float}
\usepackage{amsmath}
\usepackage{amsfonts}
\usepackage[]{url}

\usepackage{orcidlink}

\newcommand{\msun}{\mbox{M$_\odot$}}
\newcommand{\rsun}{\mbox{R$_\odot$}}

\definecolor{ochre}{rgb}{0.8, 0.47, 0.13}

\newcommand{\cmark}{\ding{51}}%
\newcommand{\xmarck}{\ding{55}}%

\title[Spinning BHs up without merging.]
{How To Spin Black Holes Up In High-Mass X-ray Binaries And Not Merge In The Attempt.}
\date{September 2021}

\author[Moreno M\'endez, Enrique.]
{Enrique Moreno M\'endez$^{1}$ 
\thanks{enriquemm@ciencias.unam.mx} \orcidlink{0000-0002-5411-9352} 
\\
$^{1}$Facultad de Ciencias, Universidad Nacional Aut{\'o}noma de M{\'e}xico, A. P. 70-543 04510, CDMX, M\'exico. 
}

\begin{document}

\maketitle
\begin{abstract}
Astrophysical black holes (BHs) can be fully described by their mass and spin.  However, producing rapidly spinning ones is extremely difficult as the stars that produce them lose most of their angular momentum before the BH is formed.  Binaries where the progenitor is paired with a low-mass star in a tight orbit can produce rapidly spinning BHs (through tides), whereas those with massive companions cannot (as they do not fit in such an orbit).
A few rapidly-spinning black holes (BHs) have been observed
paired with very massive companion stars, 
defying stellar-formation paradigm.
Models which reduce the stellar-core--envelope interaction (and winds) do not match observations nor theory well; I show they also miss explaining the energetics.
BH spins cannot be produced during stellar collapse; using orbital spin through explosion-fallback material does not match the observations; spinning BHs up through accepted mass-transfer channels takes longer than their lifetimes, it is usually discarded. 
I show that fast mass-transfer mechanisms, predicted to merge the BH and star, successfully spin the BHs up and show a mechanism to avoid said mergers and main dangers of the alternatives while naturally explaining the observations.
The implications are potentially paradigm-shifting and far-reaching in the high-energy BH-astrophysics context.
\end{abstract}

\section{Introduction: The HMXB evolution context.}
\label{sec: intro}

Cyg X-1, LMC X-1, M33 X-7, and IC10 X-1 are four high-mass X-ray binaries (HMXBs \footnote{Please refer to Table 1 for a list of acronyms and symbols used throughout this article.}; i.e., a class of BH binary, BHB, where a BH is paired with a massive companion star, $M_{\star} > 10 \msun$, with ongoing mass transfer; see Tables 2 and 3 for a list of their physical parameters); each hosting a Kerr BH (KBH; a BH with large spin or Kerr parameter, $a_\star \gtrsim 0.8$)\footnote{The Kerr parameter is defined as $a_\star \equiv cJ_{BH}/GM_{BH}^2$, where $J_{BH}$ is the angular momentum of the BH, and $M_{BH}$ its mass; $|a_\star| < 1$ for any KBH; $a_\star = 0$ represents a (non-spinning) Schwarschild BH (SBH).}.  
HMXBs are intermediate stages of systems which produce a number of high-energy transient events.  
These include Hypernovae/Gamma-Ray Bursts (HNe/GRBs; see Appendix A), short GRBs (if at least a NS is present and $q$, the mass ratio, is not far from 1), and binary BHs (BBHs) which, when they merge, are sources of the energetic gravitational-waves events (GWs) detected by LIGO-Virgo.  
They are also candidates for ultraluminous X-ray sources (ULXs; Appendices E, J).  
Their formation-channel mechanism has, so far, eluded being understood.  

There are stellar evolution models which try to explain the origin of large-spin BHs in HMXBs by reducing the Taylor-Spruit dynamo \citep[TS;][]{2002Spruit} and stellar winds from their evolution \citep[e.g.,][; but see \cite{2019Fuller} for the latest strong caveats against reducing TS; Appendices F,K]{2019Qin} and, as I will show in the next section, have a contradictory account of the formation of their KBH.  
\citet{2016MM} have shown that imparting such spins at birth is not possible with SN kicks nor standing-accretion-shock instabilities (SASIs).  
Meanwhile \citet{2017Batta,2018Schroder} have shown that angular-momentum-imbued fallback from CC SN  could spin the BHs up to $a_\star \lesssim 0.8$ \citep[again, with caveats from][see Appendix B where it is also shown that their parameters are anticorrelated to the pertaining observables]{2019Fuller}; however this would require fine tunning just to accommodate the more modest spins among these HMXBs.

\begin{table*}[H]
\hspace{-1.4cm}
\begin{tabular}{ |c|c||c|c| } 
\hline\hline
Full name&Acronym&Full name&Acroynm\\ 
\hline\hline\rowcolor{lightgray}
Black Hole&BH&Kerr parameter (BH spin)&$a_\star$\\
Schwarzschild Black Hole ($a_\star\sim0$)&SBH&Kerr Black Hole($a_\star\gtrsim0.8$)&KBH\\
\rowcolor{lightgray}
High-Mass X-ray Binary&HMXB&Intermediate-Mass X-ray Binary&IMXB\\
Supernova(s)&SN(e)&Low-Mass X-ray Binary&LMXB\\
\rowcolor{lightgray}
Core Collapse&CC&Neutron Star&NS\\
Hypernova(s)&HN(e)&Shock Wave&SW\\
\rowcolor{lightgray}
Gamma-Ray Burst&GRB&Standing Acretion Shock&SAS\\
Hypercritical Accretion&HCA&Standing Acretion Shock Instabilty&SASI\\
\rowcolor{lightgray}
Super-Eddington Accretion&SEA&Common Envelope&CE\\
White Dwarf&WD&Grazing Envelope&GE\\
\rowcolor{lightgray}
Mass Transfer&MT&Gravitational Wave&GW\\
Taylor-Spruit&TS&Black-Hole Binary (1 BH-1 star)&BHB\\
\rowcolor{lightgray}
Blandford-Znajek&BZ&Binary BH (2 BHs)&BBH\\
Blandford-McKee&BM&Ultraluminous X-ray Source&ULX\\
\rowcolor{lightgray}
Roche Lobe&RL&Roche-Lobe Overflow&RLOF\\
Laser Interferometer GW Observatory&LIGO&General Relativity&GR\\
\rowcolor{lightgray}
Proto-Neutron Star&PNS&Quantum Mechanics&QM\\
Kelvin-Helmholtz&KH&Main Sequence&MS\\
\rowcolor{lightgray}
Compact Object (BH or NS)&CO&Chemically-Homogeneous Rotationally Mixed&CHRM\\
Wolf-Rayet&WR&Bondi-Hoyle-Lyttleton&BHL\\
\hline\hline
\end{tabular}
\caption{Table of main acronyms and symbols used throughout the text.}
\label{tab:acronyms}
\end{table*}
  
Mechanisms were presented in \citet{2008MM,2011MM} which involve transferring mass from the stars and onto the BHs to spin them up but in systems which are clearly unstable under the current mass-transfer paradigm \citet{2021Klencki,1997Soberman} and would produce stellar mergers rather than the observed HMXBs.  
In this article I show that, by employing a couple of simple ideas  based on forever foregone and  overlooked aspects of hypercritical accretion \citep[HCA; see section~\ref{sec:SEA/HCA}; ][]{2017MM,2019LC,2020LC} and/or Super-Eddington accretion (SEA; section~\ref{sec:SEA/HCA}), it is possible to stop the mass-transfer episode in a common-envelope (CE; section~\ref{sec:CEsep}), and/or grazing-envelope phase \citep[GE; section~\ref{sec:CEsep}; ][]{2015Soker,2017Shiber,2018Shiber,2017MM,2019LC,2020LC} and allow the formulation of simple evolutionary tracks for the presently observed HMXB systems (Appendices G, H, Table 4).  
Henceforth, removing the need to invoke a large stellar spin, along with the risk of a HN/GRB explosion during the BH-formation process which would likely disrupt such BHBs, and prevent the formation of these HMXBs.

Discussions on the implications of these results for ultraluminous X-ray sources (ULX; by examining the likelihood of a long-lasting or a pulsed GE phase in HMXBs systems; Appendices E,J), HN/GRB central engines (Appendix A), and the implications for the formation of LIGO gravitational-wave (GW; Appendix D) sources from binary BHs (BBHs), as well as the implications for future stellar evolution models are included ; this, along with a general description of a stellar evolution scenario (Appendix D and Figure 1), in the context of our results, which outlines the formation channel which naturally explains the observables in the six known BH HMXBs with emphasis on the four with KBHs measurements and predictions for the other two.
A set of possible tests for our model can be found in Appendix G.


\begin{table*}
\begin{tabular}{ |c|ccccc|c|c| } 
\hline\hline
\rowcolor{lightgray}
(1)&(2)&(3)&(4)&(5)&(6)&(7)&(8)\\
HMXB & $M_\star$ [$\msun$] & $M_{BH}$ [$\msun$] & $q$ & P $_{orb}$ [days] & $a_\star$ & $E_{rot}$ [B] & Refs.\\
\hline\hline
\rowcolor{lightgray}
M33 X-7 & $70.0\pm6.9$ & $15.65\pm1.45$ & 10$\rightarrow$5  & 3.45 & $0.84\pm0.05$ & 3,400 & i, ii, iii \\ 
Cyg X-1 & $40^{+2}_{-2}$ & $21.2\pm2.2$ & 5$\rightarrow$2 & 5.60 & $>0.9985 (3\sigma)$ & 10,000 & iv, v, vi \\ 
\rowcolor{lightgray}
LMC X-1 & $31.79\pm3.48$ & $10.91\pm1.41$ & 7$\rightarrow$3 & 3.90917 & $0.92^{+0.05}_{-0.07}$ & 3,000 & vii, viii \\ 
IC 10 X-1 & $17-35$ & $10-15(?)$ & $2\rightarrow1$(?) & 1.45175 & $\gtrsim 0.8$ & $>1,300$ & ix, x, xi \\ 
\rowcolor{lightgray}
\hline\hline
NGC 300 X-1 & $26$(?) & $17\pm4$(?) & $?\rightarrow1.5$ & 1.366 & $>0.8$? & $>1,500$? & xii, xiii \\ 
Cyg X-3 &  $(8-15)$(?) & $\lesssim(5-10)$(?) & $?\rightarrow1.5$ & 0.20 & $\lesssim 1$? & $\sim 5,000$? & xiv \\ 
\hline\hline
\end{tabular}
\caption{Relevant properties of HMXB systems with large $a_\star$ BH: (1) HMXB name. (2) Mass of stellar companion. (3) BH mass; (4) Mass ratio (given by $q \equiv M_\star/M_{BH}$) necessary change for Conservative mass transfer onto the BH to acheive the observed $a_\star$ by HCA). (5) Shows the currently observed orbital period. (6) Lists the current measured BH spin $a_\star$. (7) Current rotational energy, $E_{rot}$, with units in Bethes, estimated from the currently observed BH spins and masses. Finally, (8) has the references for the observational data. Data with ``(?)" takes best estimates in the literature. Data with ``?" are best guesstimates in this paper from available data and extrapolating from binary evolution. The references are as follows: i, ii, iii are \citet{2008Liu,2007Orosz,2010Liu}; iv, v, vi  are \citet{2021Zhao,2003Mirabel,2021Miller-Jones,2021Kushwaha}; vii and viii are \citet{2009Orosz,2009Gou}; ix, x, xi are \citet{2015LaycockA,2015LaycockB,2016Steiner}; xii and xiii are \citet{2010Crowther,2021Binder}; and xiv is \citet{2017Koljonen} respectively.}  
\label{tab:HMXBsA}
\end{table*}

\section{The BH spin-energy conundrum.}
\label{sec:spin-E}

Let us first analyze a couple of revealing and relevant questions for HMXB systems, where KBHs lurk, in order to elucidate the main issue behind the evolutionary channel leading to their formation and why, the natal KBH hypothesis, used in most of the literature related to these HMXBs, on top of the convoluted evolutionary channels, has a fundamental flaw at its core :
\begin{itemize}
\item{A very massive star is about to collapse into a non-rotating, or Scwarschild BH (SBH).  
Will it produce a SN explosion or will it collapse directly?  
Unfortunately, the answer is not staright forward. 
Many large scale properties of the collapsing star must be considered. 
But, furthermore, SN simulations, necessary to answer such a question, are particularly intrincate computations, as they require many microscocpic, mesoscopic and macroscopic minutiae to be very well implemented, as well as GR and QM effects to be correctly quantified \citep[see, e.g.,][]{2012Janka}.  
Even though the star posseses enough energy to blow itself apart (about $10^2$ times the energy necessary to unbind all the stellar envelope) in a SN explosion following its core collapse (CC) --while its stellar core collapses and forms a compact object (CO; i.e., either a neutron star, NS, or a BH) it releases its binding energy in a fraction of a second-- the answer is not trivial because about 99\% of this energy will be emitted as neutrinos, and most of these, may leave the star without ever interacting with it \citep[see][for details]{1979Bethe}.  
Furthermore, the collapsed core (which becomes partially neutrino-opaque) forms a hot proto-NS (PNS), which radiates away great amounts of energy as it cools down and could help the explosion succeed, however, the shockwave (SW) that may give rise to the SN is in the worst place in the universe for its survival.  
The SW (supersonic by nature) stalls at a few $10^7$ cm in a supersonic inflow of the outer layers of the Fe-core of the star (photodisintegration of this Fe into neutrons and protons removes about 1 B per $0.1 \msun$ that crosses the SW\footnote{1 Bethe $\equiv 1 $ B $ \equiv 1 $ foe $ \equiv 10^{51}$ erg}), forming a standing accretion shock (SAS).  
If the SW can be revived, the SN will proceed and the PNS will cool down into a NS or a low-mass SBH.
However, if no mechanism is found to channel part of the neutrino energy into the SAS, the latter may subside and the SN will fail, allowing, thus, for the full core, as well as most of the stellar envelope to collapse into a more massive SBH (a fraction of the mass is lost through neutrinos, allowing for the release of the  gravitational binding energy); in a sense, we have the {\it unstopable force vs the unmovable wall} problem to solve and find out what kind of stellar SBH has formed, massive or not}.
\item{Now, let us compare to a collapsing a star which is allowed to have every element, every trick in the SN-explosion toolkit available --or, even one up-- in the hypernova/gamma-ray-burst (HN/GRB) central-engine toolkit to build up its chances of exploding \citep[i.e., a Collapsar;][]{1999MacFadyen}; let us as well allow it access to an energy reservoir about a hundred times larger than that for typical CCSNe, and which is readily available (as kinetic enregy) even before CC. 
Would that star explode or keep that energy stored?  
Such a question is much easier to answer!  
The star would produce as glorious an explosion as allowed and radiate a substantial amount of its available energy outright!   Right?}
\end{itemize}

Well, unfortunately, this natal-BH-spin/energy argument has not been brought up in the context of the natal spin of KBHs in HMXBs until now.  So we will address it in some depth next.

Ever since the spin parameters of BHs in HMXBs started to be estimated by different methods, and they were found to be extremely large, $a_\star \equiv Jc/GM^2 > 0.8$ (Table 2, column (6)), a long list of papers have claimed that these BH spins have to be natal \citep[from the first for M33 X-7,][to the last estimate for Cyg X-1, \citealt{2021Zhao}]{2008Liu}\footnote{This should include, basically, every article with a BH spin measurement (for HMXBs) as well as evolutionary model to explain them.}, i.e., all the BHs in HMXBs were born KBHs, along with the immense ammount of stored energy this implies.  
Now, the key point against this line of argument is this: 
Other than the large $a_{\star}$ observed in these KBHs, all available dynamical evidence points to the contrary: very low energy events formed these BHBs! 
E.g., all these systems have massive stellar BHs, so they involve either heavy fallback of a mostly failed SN or direct collapse (Cyg X-3 could be the exception here; there is evidence it could be an IMXB; where its small orbital separation suggests it could have had access to a central engine and still survived the BH formation; Appendix A), they are still bound to their companion stars (which is evidence of a low energy BH formation event, even while considering the flagrant observational bias); eccentricities are low \citep[and tidal circularizations are of order Myrs,][]{1977Zahn,1975Zahn,1989ZahnA,1989ZahnB}; have low peculiar velocities \citep[e.g., Cyg X-1 and its low velocity with respect to the Cyg OB3 cluster,][; Table 3 (column 12), which implies a $\Delta M_{SN} \lesssim 1 \msun$, and given its 2 kpc distance to OB3, a BH age of $(7\pm2)$ Myrs\footnote{These estimates might need to be updated after the latest corrections to the distance by \citet{2021Miller-Jones}.}]{2003Mirabel}; estimates of the energy of the SN (e.g. from SN remnants) that gave rise to such a KBH are, unfortunately, not available; no evidence of large jets (or BH spin orientation) to orbit missalignment (see discussion on Cyg X-1 at the end of next section), nor strong precession (from SN kicks; instead, alignmet would be expected if the spin is a product of mass transfer onto the BH); there is no evidence of SN or, better yet, HN pollution on the companions, although this is difficult to corroborate (the material will dilute into the star with time); regardless, these absences can help constraint the minimum age of the BHs.  
Kicks and mass-loss during a CCSN would, typically, have increased their previous orbital separations, so it may also be suspicious that they did not previously synchronize their spin with their orbital period; if they had done so, they would have slowed down to spins below $a_\star \lesssim 0.2$ \citep{2008MM,2011MM}. Tables 2 and 3 show the available data.

\begin{table*}
\begin{tabular}{ |ccccccc|c| } 
\hline\hline
\rowcolor{lightgray}
(9)&(10)&(11)&(12)&(13)&(14)&(15)&(16)\\
Sp.T. & $L_X$ [$10^{38}$ erg] & $e$ & $V_{pec}$ [km s$^{-1}$] & $E_{bubble}$ [erg] & SNR? & $J-S$ & Refs.\\
\hline\hline
\rowcolor{lightgray}
O(7-8)III & 0.13-2.49 & 0.0185(77) & -- & -- & -- & $\le3^{o}$ & xv, xvi, xvii, xviii \\ 
O9.7Iab & $0.05$ & 0.0189 & $9\pm2$ & $7\times10^{48}$ & Bubble & $(10-20)^o$& xix, xx, xxi, xxii \\ 
\rowcolor{lightgray}
O7III & 2.3 & 0.0256(66)? & -- & -- & -- & -- & xxiii, xxiv, xxv \\ 
WR & $0.7$ & -- & -- & -- & -- & -- & xxvi \\ 
\rowcolor{lightgray}
\hline\hline
WR & $2.6(1)$ & -- & -- & -- & -- & -- & xxvii, xviii \\ 
WN(4-7) & 0.5 & -- & $\ge 9$ & -- & Yes & ($>15^{o}$)? & xxix, xxx \\ 
\hline\hline
\end{tabular}
\caption{Spectral types and relevant dynamical properties of HMXB systems (same HMXB order as in (1), Table 1) with large $a_\star$ BH: (1) HMXB name. (9) Spectral type (Sp. T.) of stellar companion. (10) X-ray luminosity of the BHB; (11) Eccentricity of the binary. (12) Peculiar velocity vs cluster. (13) Energy deposited into bubble by jets. (14) Does it have an observable SNR? (15) Orbital to spin (jet) inclination. (16) References for the observational data. The references are as follows: xv, xvi, xvii, xviii are \citet{2008Liu,2007Orosz,2010Valsecchi,2010Liu}; xix, xx, xxi, xxii are \citet{2021Zhao,2021Kushwaha,2005Gallo,2014Casares}; xxiii, xixv, xxv are \citet{2009Orosz,2009Gou,2014Casares}; xxvi is \citet{2008Barnard}; xxvii and xxviii are \citet{2008Barnard,2015Binder}; xxix and xxx are \citet{2009Sakari,1996vanKerkwijk}, respectively.}  
\label{tab:HMXBsB}
\end{table*}

In order to estimate the available energy available to form a BH of mass $M_{BH}$ and spin $a_\star$ we can use (e.g., \cite{2011MMB}):
\begin{eqnarray}
E_{KBH,rot}& = &f(a_\star) M_{BH} c^2, \;\;\;\;{\rm where} \nonumber \\
f(a_\star)& = &1 - \sqrt{\frac{1}{2}\left(1+\sqrt{1-a_\star^2}\right)}
\end{eqnarray}
to obtain the order of magnitude of available kinetic energies during and after CC if these BH spins were natal, as claimed (estimates listed in Table 2, column (7)).  Having KBHs with a natal spin of such magnitude and such large BH masses, $M_{BH} > 10 \msun$ for all for well measured HMXBs (Table 2, columns (3) and (6)), implies that they had between $10^{54}$ and $10^{55}$ ergs available during CC!  Even though, the available energy to a KBH central engine will be a factor of 3 to 4 below the rotational energies listed, all the kinetic energy is fully available to the star and its collapse dynamics before the BH is formed \footnote{This is also a lower limit, considering for efficiencies and an actual energetic explosion which still left such a KBH behind.}, so accretion disks and extremely strong magnetic fields will be anything but absent. 
Thus, the energy to blow themselves appart, by forming a centrifugally supported accretion disk, and building up to $10^{15}$-G magnetic fields encompasing some $V \sim (10^{7} {\rm cm})^3$ wide volume is present (they may be one or two orders of magnitude less intense as well as one order of magnitude less extended, but its enlightening to look at the {\it comparatively low energy cost} to produce them, for some $10^{51}$ erg, when compared to the energy budget proposed by such an evolutionary model); the formation of such KBH should have produced some spectacularly luminous events, a large amount of mass loss, and disrupted the BHBs, or at least, left very strong imprints in them which should be visible today, only, {\it none of them did!} (see Appendix H for further discussion on this topic).  

In principle, one could attack the energy argument by saying that these KBH have so much energy that they rapidly dismantle the central engine and stop the energy drain, keeping the KBH almost intact \citep[the Goldielocks scenario in][]{2011MMB,2007Brown,2008Brown,2014MM}, but this alternative can easily be discarded, for this scenario, given that the (kinetic) energy is present before CC and can find different paths to be used to reduce the natal mass and spin of the BH before it is even formed.
Appendix A, in the supplementary sections, addresses isues on the formation of KBH central engines of HN/GRBs (where the BH spin {\it must} be natal) and its relation to X-ray binaries (XBs).  In general, it is more likely HNe/GRBs come from IMXB, rather than HMXBs (for the formation of the first BH; the formation of the second BH could produce HNe/GRBs in either case).

One model which cannot be dismissed by the energy argument is that the spin may be acquired by caputring orbital angular momentum through SN-fallback material \citep{2017Batta,2018Schroder}, nonetheless, as shown in Appendix B, it has a number of important caveats to overcome.

Fortunately, there is an alternative explanation: HCA during CE or GE evolution after the formation of the BH.  Nonetheless, it involves accepting that a couple of dogmas have to be thrown out the window; I will show in the following section how these may be easily overcome. 
I will now provide a channel that prevents the merger during this stage, and naturally explains other properties observed in these HMXBs.


\section{On the Eddington limit, BHL, HCA/SEA.}
\label{sec:SEA/HCA}

When CE evolution is studied in a binary stellar system, the usual set up for studying the removal of the CE and inspiral of the orbiting stars uses an energy-conservation prescription (e.g., \cite{2021Klencki,1997Soberman,1994Shore}):
Gravitational binding energy of the CE vs the energy difference between initial and final orbits.
Usually mediated by the efficiency by which the energy is converted ($\alpha$) and a radial density profile of the stellar material in the CE ($\lambda$), which can be treated as a product ($\alpha\lambda$), and binary systems which have undergone a CE phase may be used to constrain/calibrate it \citep[see ][for XR binaries, where they find $\alpha\lambda \sim 0.2$]{2002Lee}.

Nonetheless, all HMXBs in our study, Cyg X-1, LMC X-1, M33 X-7, and IC 10 X-1 have mass ratio between companion star and BH ($q_{\rm system} \equiv M_{\star}/M_{\rm BH}$) well above 1 (Table 1, column (4)), hence RLOF, CE/GE mass transfer will be driven by at least one runaway process i.e., once it begins, it cannot be stopped until a merger occurs
, unless $q$ can be brought below $q \lesssim 1$ beforehand.  The actual state of these binaries, then should \underline{not} be attainable via a RLOF or CE where the donor star transfers mass onto the BH, instead; they should be either in an ongoing CE stage or they should have merged if RLOF has ever ensued after BH formation. In principle this argument should discard the HCA during CE/GE scenario for BH spining in these binaries from \cite{2008MM,2011MM}, but see section~\ref{sec:model}.

Unfortunately, most of the research in this field invokes the Eddington limit (for short reviews on the topic, see e.g., \cite{1994Brown,1989Chevalier}) for Bondi accretion onto the BHs (i.e., spherically symmetric; \cite{1952Bondi}) which can be obtained from:
\begin{eqnarray}
\dot{M}_{\rm Edd} &\simeq& \frac{L_{\rm Edd}}{c^2} \simeq 10^{-15} M_{10}~\msun ~{\rm s}^{-1}, \; {\rm and} \nonumber \\ 
L_{\rm Edd} &\simeq& 10^{39} M_{10}~{\rm erg~s}^{-1},
\label{Edd}
\end{eqnarray}
($M_{10}$ stands for a unitless $10-\msun$ accretor)
to argue that, given the masses of the companions in these systems (Table 1, (2)), their ages constrained to $\tau_{\rm HMXB} \lesssim 10$ Myr (even after accounting for stellar rejuvenation of the companion stars from mass transfer from the BH-progenitor stars \cite{1995Braun}), the requirement of transfering $5~\msun$ to $15~\msun$ onto the BHs (to spin them up to $a_\star > 0.8$ from initial low values) at the Eddington rate is not attainable (\cite{2019Qin,2019Fuller,2008Liu,2021Zhao,2014McClintock,2010Liu,2009Gou,2016Steiner,2010Valsecchi} and a large etc.) as more than 10 Myrs are necessary.  

However, these systems, either during RLOF mass transfer (or wRLOF; \cite{2007Mohamed,2018ElMellah,2019ElMellah}), or during CE mass transfer, are under conditions which are extremely different from Bondi accretion.  In any case, Bondi-Hoyle-Lyttleton \cite{1944BondiHoyle,1941HoyleLyttleton,2004Edgar} accretion is a much better approximation to what is occurring, and as such, it should be considered in such estimates. Still, BHL falls short from considering the density, velocity, and temperature gradients, as well as the Coriolis and Centripetal/Centrifugal forces present during the CE stage, even after jets are considered; numerical simulations have been performed in order to account for these in \cite{2017MM,2019LC,2020LC} (where even hints of accretion-disk-like flows are observed
) and can be used to constraint the near BHL accretion rates for BH accretion during CE (finding typical values of $\dot{M} \simeq \eta \dot{M}_{\rm BHL}$ with $\eta \sim 0.1$).  Along with the substantial amount of transferable material, these conspire against the Eddington limit and allow for mass-transfer rates which not only are super Eddington\footnote{$\dot{M}_{\rm Edd} < \dot{M} \lesssim 3,000 \dot{M}_{\rm Edd}$; mostly due to the geometry of the flows: equatorial accretion, poloidal radiation or jets, thus avoiding the main Bondi accretion bottleneck.}, but even hypercritical (where cooling is mostly done through neutrino transport \cite{1994Brown,1989Chevalier}). In fact, accretion rates have been estimated numerically as well in the simple BHL mode for the newtonian case see, e.g., \cite{1994Ruffert}; for the GR, or higher resolution of the accreting BH case, convergence is quite clear \citep[see][this result includes density gradients]{2015Lora-Clavijo}.

So, we have now shown that the Eddington luminosity is nowhere near an impediment to accrete the necessary mass to bring a BH from $a_\star$ from close to 0 all the way to 1 within a small fraction of the lifetime, $\tau_{\rm HMXB}$, of these binaries; i.e., doubling the initial mass of the BH (see Fig 6 in \cite{2000Brown}, as suggested in \cite{2008MM,2011MM}) within a much shorter period of time as shown in \cite{2017MM,2019LC,2020LC,2022MM} where a CO within the CE of a RG or RSG accretes at about $\eta \dot{M}_{\rm BHL}$:
\begin{eqnarray}
    \dot{M}_{\rm BHL} \equiv \epsilon \frac{4 \pi \rho G^2 M^2}{(v^2+c_s^2)^{3/2}} 
    &\simeq& 10^{-5} \ \epsilon_{0.1} \ \rho_{-6} \ M_{10}^2\ v_7^{-3} \ {\rm M}_\odot \; {\rm s}^{-1} \nonumber \\
    &\simeq& 10^{10} \dot{M}_{Edd}.
  \label{eq:BHL}
\end{eqnarray}  
Thus, a BH of $5~\msun$ may double its initial mass within a month and a $10~\msun$ BH in a few days if the mass is available (i.e., within its RL); even at accretion rates well below those of BHL, e.g., for SEA $\dot{M}_{\rm SEA} \lesssim 10^{3-4} \dot{M}_{\rm Edd}$, the necessary mass transfer and accretion can easily be acheived within a Myr.

\section{On CE separation by ablation and by accretion.}
\label{sec:CEsep}

From section~\ref{sec:SEA/HCA}, we see that the elephant in the room is not whether a BH can accrete enough mass (which it shows can be done with HCA/SEA), if available from a massive companion, but rather, how to stop the BH from merging with and swallowing its companion star given that the mass transfer will be unstable until the BH has more mass than the donor (see \cite{2021Klencki,1997Soberman,1994Shore} for a more in depth analysis), which is not close to the situation observed in most of these HMXBs.

Happily, the issue can be resolved by considering two tools available to HCA/SEA due to the BHL accretion rate.
The first one, a usually (and correctly) disregarded issue in most CE binaries studies but which is highly relevant here; it was suggested for a GE of a NS and a RSG in \cite{2015Papish}, further studied (NS/BH with RSG) in \cite{2017MM,2019LC,2020LC}; here I propose it works in much more bound stellar environments as well, as detailed next.  An accreting CO, either a NS or a BH, has a very deep potential well, unlike most stars, and thus, the binding energy released by such accretion is of order 
\begin{equation}
E_{\rm B} = \epsilon M_{\rm acc} c^2
\label{EBind}
\end{equation}
with typical $\epsilon \sim 0.06$ to $\sim 0.1$ for NSs and SBHs ($a_\star \sim 0$) and as large as $\sim 0.4$ for extreme KBHs ($a_\star \lesssim 1$).  This translates to a binding energy ($E_{\rm B}$) release of order $10^{54}$ erg to $7\times10^{54}$ erg when the accreted mass is of order $M_{\rm acc}\sim 10 \msun$.  But these are energies comparable to those available to the central engine of a GRB/HN event (of course, the luminosities are much smaller, being determined by the accretion rate).  Even if 99\% of the energy is emmitted as neutrinos this is more than enough energy for jets \citep{2017MM,2019LC,2020LC,2015Soker,2017Shiber,2018Shiber,2015Papish} to remove a $M_{\rm CE}\gtrsim 40~\msun$ CE (where I take the stellar radius to be $\sim 4 \rsun$; these are extreme values, but they make the point that they work even for M33 X-7, the most massive HMXB in our sample) with a binding energy of $E_{\rm B,CE} \lesssim 10^{51}$ erg. This can be achieved (in as little as a few days to as long as a few thousand years) with BHL SEA/HCA (i.e., accounting for 1$\%$ neutrino- and 10$\%$ BHL-efficiency, for a total $0.1\%$ efficiency).
In escence, this means that, a 10-$\msun$ BH which doubles its mass by BHL accretion will release enough energy to unbind the CE/GE in which it participates. 
Hence, the mechanism proposed in \cite{2008MM,2011MM} can explain the formation mechanism for the HMXBs with high-spin BHs without incurring into an early merger; they may be born SBHs and HCA until the BH masses and spins have reached their current values!  
This, of course,  has the observational advantage that the natal BH masses can be large ($M_{\rm BH} \gtrsim 10~\msun$, allowing the binary to survive the formation of the BH), and the SNe were, at best, low energy, if not failed, thus in agreement with the large BH masses observed in the HMXBs. The substantial amount of mass loss through this channel may prevent the orbital separation to decrease.  It is important to notice that this mechanism is self-regulated, if too much energy is produced by, e.g., jets, accretion rates drop and so does the jet luminosity, thus allowing the accretion to increase again \cite{2019LC}, so until, enough accretion and released energy have not been acheived, the CE will not be lifted.

We may call such a mechanism {\it CE-separation by ablation}, and note that, under the right conditions, it may lead to an extended period of a GE \cite{2015Soker,2017Shiber,2018Shiber} or pulsed GE (which we may define as when the mass transfer is intermitent due to KH; i.e., if the stellar envelope is convective \cite{2021Klencki,1997Soberman,1994Shore}).

As mentioned above \cite{2015Papish}, under the assumption that accretion disks form and launch jets, for systems involving WD, MS, and AGB stars, jets can remove a CE; here, we have borrowed the model for HMXBs and use the larger energy deposition of HCA/SAE onto the BHs, plus we note that jets can do the job as shown in \cite{2017MM,2019LC,2020LC} (for a CO-RG CE; and for a MS in our previous estimate), especially if an accretion disk can be formed, which is likely in a GE, but it is not always clear within a CE \citep[see][]{2022LC}, however, any photon or kinetic outflows which may deposit the large amount of energy by the release of the gravitational binding energy may do the job well enough \cite{2019Aguayo,2021Aguayo}; thus, to allow for larger photon luminosities, accretion rates close to the HCA/SEA boundary would probably be favored.

For SEA (given its nature, a mostly geometrical effect) to occur it is important to have something similar to BHL accretion, such that an accretion disk forms (or a configuration that allows equatorial inflows and polar outflows); such conditions naturally arise during RLOF, wRLOF (\cite{2007Mohamed,2018ElMellah,2019ElMellah}\footnote{This is a very relevant mode in these HMXBs \citep{2018ElMellah,2019ElMellah}, as it is close to current observed status in the six HMXBs we discuss in this paper; see the current x-ray luminosities in column (10) of table 3, they are consistent with wRLOF MT at, e.g., $\sim 10^{-5} \msun$ yr$^{-1} \sim 10^{20}$ g s$^{-1}$, for an accretion efficiency of $\sim 10\%$, and $1\%$ in photons, this gives $L\sim10^{38} erg$ s$^{-1}$.}, 
GEs, and/or CEs and that is why these stages in binary stellar evolution should be more carefully considered. 
BHL facilitates equatorial accretion (mostly driven by angular momentum transfer) and polar outflows, e.g., photon radiation (jets).
For HCA, independently of the geometric components\footnote{E.g., CCSNe are HCA by nature, but the accretion is spherically-symmetrtic, i.e. Bondi, however, due to the extreme conditions of the infalling material, the Eddington limit is that for neutrinos (instead of photons), which as mentioned earlier, help support the SAS until the Fe core has fully collapsed and the SW may be revived.}, neutrino processes alleviate the radiation transport of the binding energy due to an increase in accretion rate boyond the limit of SEA.  So, the issue of how much energy is radiated by photons and how much by neutrinos arises (CEs are transparent to neutrinos, so they cannot help remove the CE).  
At the HCA/SEA limit we have $\sim 100\%$ photons, whereas at CC we have 99\% neutrinos, so we can always pick an efficiency between 1 and 100\% which will introduce an error bar of a couple of orders of magnitude in our energy budget estimates.  
For 10\% of BHL HCA onto a 10-$\msun$ BH (eq.3, section~\ref{sec:SEA/HCA}), this amounts to a (photon) jet luminosity range of $L_{\rm jet} \lesssim 4\cdot 10^{(47-49)}$ erg s$^{-1}$.  Which means, that in a matter of a few hours to a few days, it could remove the CE of a massive star such as those in HMXBs such as M33 X-7, LMC X-1, or Cyg X-1.  
A CE lasting up to a few $10^3$ years could be removed with a jet of $L_{\rm jet} \gtrsim 10^{(40-41)}$ erg s$^{-1}$, below that, a merger stage could be initiated.
Of course, this scenario should leave around a large amount of energized ablated material that could be currently observable.

An immediate second tool, so simple and elegant it was hidden in plain sight, now appears:  What if all of the CE-available material is accreted onto the BH?  
After removal of the binding energy and angular momentum by neutrinos, which do not lift away the CE material, this option is quite likely for the BHL accretion regime (and, obviously, overlooked and never considered in the Eddington-limited-accretion-rate regime); it was even part of the CC process during BH formation!  
But wait, is that not a merger by definition?
Notice that this is what occurs during a $q$-runaway, on a dynamical timescale.  
True, but as long as $\dot{M}_{\rm BHL} > \dot{M}_{\rm MT}$ everything is solved when $q \rightarrow 1$, as the $q$-runaway will stop.  
The $\alpha\lambda$ CE is no longer an issue either as all material is depleted by accretion onto the BH as well. 
It also solves part of the problem when the jets are not luminous enough to lift the CE and only the KH-runaway is active. 
In the HCA regime (B or BHL), all matter that can get rid of its excess angular momentum and kinetic energy is up for grabs by the BH as long as it crosses the L1 (Lagrange 1) point, the accretion rate is a no-issue (at about $\dot{M}_{\rm BHL} \simeq 1 \msun$ day$^{-1}$ for a 10-$\msun$ BH and a $\tau_{\rm KH} \gtrsim 10^4$ yrs for the stars in HMXBs).  
As shown in \citet{2021Klencki,1997Soberman,1994Shore}, if the mass ratio reaches either $q \lesssim 1$ values (which would be unlikely before CC SN, or substantial  accretion and ablation), or if the mass transfer reaches a radiative layer in the stellar envelope, the runaway RLOF or GE/CE MT process should stop (until the stellar structure reaches a new thermal equilibrium configuration in a $\tau_{\rm KH}$ timescale for HMXBs in the former case), breaking the unstable MT and the CE phase appart; this may lead to a GE, or to a KH-mediated, pulsed-GE stage.  

One must be careful to remember that, even if the CE is radiative, but $q>1$, conservative MT keeps shrinking the orbit in a dynamical runaway; so unless MT is not conservative, e.g., if ablation helps to partially lift the CE, the merger is not prevented until $q\sim 1$; and if it does, in a radiative layer, even an underway merger should stop whenever the accretion rate surpases the RLOF-MT rate (as the $q$-runaway process is gone, and along with it, the $\alpha\lambda$ induced orbital drag/friction)!
In the convective envelope case, the second runaway will be activated on top of the $q$ runaway, and material dumped onto the BH will accrete away at nearly BHL rate in a short CE phase, which will, nonetheless, substantially alter the binary mass, spin, and orbital parameters (shrinking the orbit for $q \gtrsim 1$ and conservative MT), as soon as a radiative layer is found, or if jets are active, the merging process stops.  We may call this scenario {\it CE-separation by accretion}.   

It is suggestive to note that for the HMXBs with WRs in our sample, even if the masses are not well constrained, the $q$ values do not necessarily stray away from $q \sim 1$.  So, in principle they may have already evolved past such a {\it merging} CE/GE phase, where the primary MS underwent MT RLOF, attempted a merger, but then $q \rightarrow 1$, probably driven with CE-separation by accretion, which allowed for nearly conservative MT, and thus, the orbital period decreased from a few days to under 1.5 days, the H envelope was accreted by the BH (spining it up), and then the, already underway, so-called merger failed, providing us with further observational hints of this scenario.  In a sense, this, change of paridgm in how we approach accretion during CE MT begs the question:  Could you actually merge such systems with a BH when HCA is such a powerful available accretion tool?  It seems that, once $q \sim 1$, the answer is no. 

\section{Recipee for preventing mergers during CE in HMXBs.}
\label{sec:model}

Four mechanisms conspire to cause a merger during a RLOF or CE/GE mass transfer phase in HMXBs when the mass ratio of the star to the BH is $q \equiv M_\star/M_{BH} > 1$:
\begin{itemize}
\item The {\it $q$-runaway} process:  When the donor is more massive than the accretor, $q > 1$, conservative MT shrinks the orbital separation dynamically with swift and catastrophic consequences (the star no longer fits in its RL).
\item The {\it Kelvin-Helmholtz (KH) runaway} process but only for convective envelope (or incumbent layers of) stars or CEs. This is due to the layer being adiabatically dominated, the Mass-Radio relationship becomes inverted and the star expands when it losses mass \cite{2021Klencki,1997Soberman,1994Shore}.
\item Once the CE sets in, the $\alpha\lambda$ mechanism converts the orbit in vacuum into an orbit in a medium, and thus, friction or drag causes the orbital separation to decrease.
\end{itemize}
So, breaking links in this catastrophic chain of events can help stop mergers or at least help slow them down until another link breaks apart; of particular importance is bringing $q \lesssim 1$ which works on a dynamical timescale and precipitates the $\alpha\lambda$ mechanism.

Two important mechanisms are available when HCA/SEA appears in the scene which help break links of the merger chain of events:   
\begin{itemize}
\item Separation by ablation (section~\ref{sec:SEA/HCA} and~\ref{sec:CEsep}):  The extreme amount of gravitational binding energy released by the massive accretion process onto a CO can be used to drive jets and winds (Appendix C,H) or other outflows (e.g., cocoons) which help remove the CE and, in some cases, keep the stellar envelope within its RL \citep[see, e.g.,][]{2015Soker,2017Shiber,2018Shiber,2017MM,2019LC,2020LC}.
\item Separation by accretion (Appendices C,D):  This is due to the large HCA/SEA rate during BHL accretion.  All RLOF MT material can be accreted by the CO as fast as it is transferred due to the BHL HCA/SEA nature of the accretion flows (regulated by how fast the excess angular momentum, AM, can be lost).  Thus, this process can regulate the KH runaway, but not the $q$ runaway 
Nonetheless, this HCA/SEA accretion mechanism can rapidly (a matter of days; see eq.~\ref{eq:BHL} in section~\ref{sec:SEA/HCA}) and fully stop a merger in process by bringing $q \rightarrow 1$ and, by full CE (accretion) removal, stop the $\alpha\lambda$ mechanism.
\end{itemize}

\begin{figure*}
\centering
   \begin{tabular}{@{}c@{\hspace{.5cm}}c@{}}
       \includegraphics[page=1,width=0.7\textwidth]{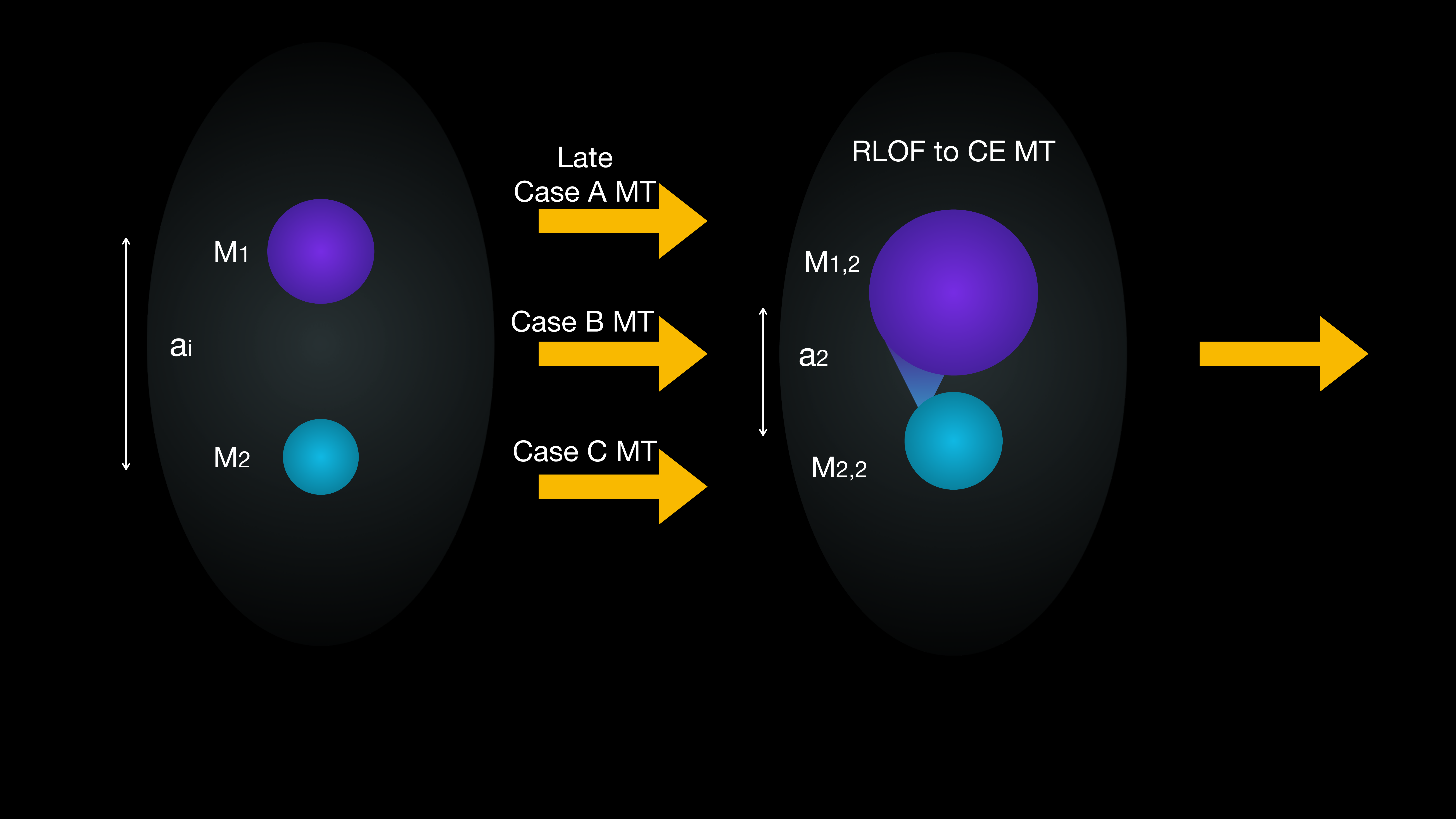} \\
       \includegraphics[page=2,width=0.7\textwidth]{HMXBevolChannel2.pdf} \\
       \includegraphics[page=3,width=0.7\textwidth]{HMXBevolChannel2.pdf} \\
   \end{tabular}
\caption{General scheme of the evolutionary channels with unstable CE stoped by ablation or accretion with HCA/SEA for HMXBs with KBHs: (1) Top-left panel: ZAMS masses $M_1 > M_2$, $a_i \gtrsim (10^{2-3} \rsun$. (2) Top-right panel: $M_1$ RLOFs, goes into a CE and transfers mass onto $M_2$, orbital separtion  shrinks, $a_2 < a_i$ to a few $10^1 \rsun$; $q$ is inverted ending the  first CE. (3) Middle-left panel: $M_{1}$ evolves into pre-CC star (orbit too wide for synchronization, else, too wide to produce large $a_\star > 0.2$ of pre-CC star), $a_2 \sim a_3$. (4) Middle-right panel: CC BH formation, mass loss to explosion makes orbit expand $a_4 > a_3$, SBH born with $a_\star > 0.2$. (5) Bottom panel: CE/GE/pulsed GE phase substantially increases BH mass $M_{BH,5} \simeq (1.5 - 2.5) M_{BH}$ to bring $a_\star$ to the 0.8 to 1 range; this mass-transfer phase is regulated/stabilized by ablation, HCA/SEA full envelope accretion, or both; ULX or pulsed ULX phase is likely in process.  A combination of masses and Cases (A, B, or C) can accomodate for the different HMXBs in Table 1.}
\end{figure*}

Finally, we can present a stellar evolution model which may connect the six observed HMXBs we present in this paper.  Particularly, explaining the difference between the O-star and the WR systems.
We will begin at the fourth panel of Figure 1.  A SBH just formed in a HMXB via a dark CC SN or direct collapse which produced a massive BH with little binary orbital perturbation.  The O-star companion has recently (in panel 2) been recycled by the CE MT which deposited the H envelope of the primary.  wRLOF ensues for a few Myrs (in the more massive HMXBs) to tens of Myrs (in the case of the lower mass systems), thus the mass and AM accretion onto the SBH begins.  
As the O-star components evolve, they begin filling their RLs, stable MT ensues due to the $q > 1$ and KH runaways.
However, these are rapidly stopped by CE-separation by ablation.  
Short episodes of RLOF MT and pulsed-GE evolution follow (thus brief ULX-like stages may be present).  
Nonetheless, these are probably resposible for most of the MT as well as the AM transfer which has produced the spins currently observed in the KBHs in these HMXBs, particularly in the O-star systems; this stage, represented in the fifth panel of Fig.1, is the current state of M33 X-7, Cyg X-1 and LMC X-1. 

In the case of the WR systems, IC 10 X-1, NGC 300 X-1 and, perhaps Cyg X-3, they may have evolved beyond.  The O-star left the MS and attempted to enter the RG, only to quickly fill both RLs (its own and that of its BH companion).  Hence, a merger scenario set in, too much MT too fast; if the BH was buried too deep into the CE, it may happen that no accretion disk nor jets are produced, thus setting up the stage for CE-separation by accretion.  The consequence being substantial conservative MT onto the BH through HCA, further decrasing the orbital separation, and bringing $q \rightarrow 1$ by enhancing the mass of the BH and removing the H envelope from the O star; thus, causing the merger to fail.  Therefore, producing the WR systems we observe today, with large BH masses, small orbital periods, and $q \simeq 1$.  Ready and prime for their debut as HN/GRB progenitors and, if they survive the fireworks, GW sources when the two KBHs merge.  It is noteworthy that HN/GRB explosions will likely disalign the spins of the KBHs with respect to each other as well as to the orbital AM (which this evolutionary channel would have previously aligned).

What factors may influence whether CE-separation by ablation or by accretion are the main channel at one point or another?
It is likely that SEA has larger photon luminosity and accretion-disk winds than HCA (which has a larger overall luminosity, but neutrino dominated), so it will probably help the ablation scenario.  Thus, SEA needs accretion disk formation in order to produce jets, whereas HCA does not.  Hence, situations where the orbital separation to stellar radius is large (small RL filling factor by the star) and the RLOF occurs rather fast ($\tau_{\rm sync} > \tau_{\rm \star,expansion}$) the stellar envelope will present itself as a large-velocity wind to the BH, along with velocity and density gradients, Coriolis, etc., thus allowing for accretion disk formation. So, important factors to consider are the number of runaway processes active during RLOF, $q$, KH, $\alpha\lambda$.  The faster the RLOF occurs the less likely tidal synchronization will be present. This would then favour CE separation by ablation.
Instead, if the star has tidally synchronized, then accretion while still assymetric, will have less AM to form an accretion disk, so jets are less likely (but winds or other polar outflows may still be present; section~\ref{sec:CEsep}), thus going into HCA rather than SEA. 
Hence, CE separation by accretion will be more likely.  

The final relevant timescale in this process is, of course, the viscous timescale required to fully accrete an accretion disk (if and when it forms) into the BH (in a sense, it allows us to estimate  how rapidly the disk/accretor/infalling matter can radiate away all the binding energy and angular momentum necessary to fall into the BH).  
Given the extended disks that may form in HMXBs (from $R_{disk} \sim 10^{12}~{\rm cm}\sim r_{RL}$, i.e., if the radius of the disk is that of the RL of the BH, and down to $10^{6}$ cm at the ISCO\footnote{Located between the {\it ms} and the {\it mb}, or marginally stable and marginally bound orbits for SBH or KBH which are located between $1 M \le R_{isco} \le 9 M$.}) this is a relevant issue.  
Using eq.~\ref{eq:BHL} in section~\ref{sec:SEA/HCA}, we see that at $\dot{M}_{\rm BHL} \simeq 10^{-5} \msun$ s$^{-1}$ and assuming that $10~\%$ of the energy is radiated (binding energy) away (e.g., by jets), then $L_{\rm BHL} \simeq 10^{48}$ erg s$^{-1}$.  
If we assume these objects are observed as extreme ULXs, and these are obesreved up to $10^{42 - 44}$ erg s$^{-1}$, we could extrapolate a maximum accretion rate between of $10^{-4}\dot{M}_{\rm BHL} \gtrsim \dot{M}_{\rm acc} \gtrsim 10^{-6}\dot{M}_{\rm BHL}$ (or $10^{6}\dot{M}_{\rm Edd} \gtrsim \dot{M}_{\rm acc} \gtrsim 10^{4}\dot{M}_{\rm Edd}$) regulated by accretion-disk viscosity.
To avoid further deviating from our objective, we can simply state that the presence of objects like ULXs to CCSNe (including all high-energy transients in between) account for the presence of both SAE and HCA in nature.  

One last caveat: \citet{2014McClintock} mention that a spin-orbit missalignment larger than $\theta_{\rm so} \gtrsim 10^{o}$ could modify the continuum-fitting $a_\star$ meassurements; Cyg X-1 has a $\sim 150$ day modulation which has been associated \citep{2007Ibramigov} to the precession of the jets (and, likely, BH-spin and inner-disk angular momentum vectors) with an orbital inclination of $(10-20)^o$. But it is unreasonable to expect a correction from $a_\star \gtrsim 0.99$ values to $a_\star \lesssim 0.2$  \citep[rather than a small decrease in total spin; see concerns expressed in][]{2019Fuller}.  
But this raises the, highly unlikely, alternative that all HMXBs {\it could} have tilted spin-orbit alignments and could be biased towards higher values of $a_\star$.  
However, there are five out of five (Cyg X-1 among them) spins which are consistent with Fe K$\alpha$ line meassurements; both these methods seem to be consistent with each other, thus strengthening the case for the large $a_\star$ measured values.  
If confirmed, the spin-orbit tilt in Cyg X-1 can help constraint the SN kick perpendicular to the pre-CC orbital plane and, thus, the explosion energy.



\section{Conclusions}
\label{sec:concl}

\begin{table*}
\hspace{-0.5cm}
\begin{tabular}{ |c|ccccccccccc| } 
\hline\hline
\rowcolor{lightgray}
(1)&(2)&(3)&(4)&(5)&(6)&(7)&(8)&(9)&(10)&(11)&(12)\\
Models & Bound & Merge & $M_{BH}$ & $M_{comp}$ & $P$ & e & $a_\star$ & $E_{rot}$ & SO align  & Bubble & Polluted\\
\hline\hline
\rowcolor{lightgray}
1& \cmark & \xmarck & \cmark & \cmark & \cmark ? & \cmark & ? & -- & -- & ? & \cmark \\
2& \cmark & \xmarck & -- & -- & -- & -- & \xmarck & \xmarck & -- & -- & -- \\
\rowcolor{lightgray}
3& \cmark & \xmarck? & \cmark? & \cmark? & ? & \xmarck & \cmark? & \cmark & \cmark & \cmark & \cmark\cmark \\
4& \cmark & \cmark & \cmark? & \cmark? & \xmarck & \xmarck & \cmark & \cmark & \xmarck & \xmarck & -- \\
\rowcolor{lightgray}
5& ? & \xmarck & \xmarck & \cmark & \xmarck & \xmarck & \cmark? & \xmarck & \xmarck & \cmark? & \cmark \\
\rowcolor{green}
6& \cmark & \xmarck & \cmark & \cmark & \cmark & \cmark & \cmark & \cmark & \cmark & \cmark & \xmarck \\
\hline \hline \rowcolor{lightgray}
Obs.& \cmark & \xmarck & \cmark & \cmark & \cmark & \cmark & \cmark & \cmark & \cmark & \cmark & \xmarck \\
\hline\hline
\end{tabular}
\caption{Summary of best estimates of different evolution model outcomes compared to observations in HMXBs (last row):
Model 1, [2]; Model 2, SASI [4]; Model 3, [5,6]; Model 4,  [7,8] under the assumption of a merger from CE; 
Model 5, [7,8]+TW (TW stands for conclusions from this work, i.e., adding the energy necessary to explain a natal $a_\star = 0$ vs a $a_\star \gtrsim 0.8$);
Model 6, [2]+TW, Final row is for the obervables in the six HMXBs considered in this work. (see Appendix C for more information).
\cmark ~stands for an observationally confirmed or positive parameter in the model; \xmarck ~for a negative result, i.e., not confirmed by observations; $-$ appears when the model does not account for the parameter; \cmark\cmark ~when it is expected to be large, and a question mark is placed next to an extrapolated parameter or guesstimate, or alone if no extrapolation is well suited or it is in doubt. The parameters answer the next list of questions/parameters: (1) Model, (2) will the system remain bound after a HN/GRB using the energy required to produce the predicted BH spin?, (3) will the system merge in a CE before producing a HMXB?, (4) Is the expected BH mass in agreement with those in the HMXB sample?, (5) Is the expected companion mass in agreement with those in the HMXB sample?, (6) Is the expected orbital period in agreement with the observed ones?, (7) How about the expected eccentricity?, (8) Is the expected BH spin observed?, (9) Is $E_{rot}$, the available energy at CC, consistent with observations?, (10) Is the expected Spin-Orbit alignment consistent with observations?, (11) Are they likely to form a accetion/jet powerwed bubble after BH formation?, (12) Is the companion heavily polluted by a SN or HN/GRB?}  
\label{tab:ModelsVsChallenges}
\end{table*}

In this article  we have shown that the formation of natal KBHs in HMXBs is not energetically consistent with the observables in these BHBs (App.A). As an aside, the conclusions in \cite{2008MM,2011MM} about the impossibilty of producing large natal spins in massive stellar BHs hold well to the observational evidence (see discussions in Apps.A,D); the stellar evolution models proposed there are now on solid ground with our novel MT model. Even after almost 5,000 stellar models in \cite{2019Qin} the natal spins are produced under assumptions which observations do not favour, as the results in \cite{2019Fuller} and other sources have shown (App.F). The alternative of failed SN fallback is still possible, but not observationally favored; it also has trouble explaining the fastest spinning KBHs (Appendix B).

I have shown that MT with HCA/SEA in CE/GE/pulsed-GE phases (KH regulated), after the formation of a SBH in HMXBs, can be stabilized and mergers after RLOF are fully preventable through the CE-separation by ablation, CE-separation by accretion, or by a combination of both; the mechanisms described in sections~\ref{sec:spin-E} and~\ref{sec:model}.
The current population of three O-star and three WR-HMXBs is suggestive of pre- and post-H-stripping during the attempted merger phases through RLOF-MT going into CE/GE stages (as in section~\ref{sec:CEsep}).
This mechanism naturally explains large BH masses, large companion masses, large BH spins, current spin-orbit alignments, low eccentricities, low peculiar velocities, low companion pollution, observed X-ray luminosities  (through wRLOF with possible intermitent periods of ULX; Appendices E,G,H,J), as well as the identified energy bubbles (which could be a SNR, or, more likely, powerwed by SEA, low HCA periods, or mass ejected by CE/GE phases); we have discussed most of these parameters, currently observed in some of the HMXBs.  
Orbital periods, efective temperatures, may be easily fitted as well (see Table 4 along with its explanation in Appendix E). Appendix G has a set of possible tests for some of the hypotheses and results put forward in this article.

I have discussed possible outlines for the evolution giving rise to these HMXBs utilizing the new tools at the disposal of HCA or SEA (more binary stellar evolution details can be found in Appendices G,H).  

I have also provided highlights of the consequences of our scenarios for observational consequences for HMXB evolution, ULX sources, the production of HN/GRB central engines (see Apps.A,E) as well as the production of LIGO sources, which, even accounting for possible HN/GRB events, may be aided at remaining bound by a possible BH-mass distribution bifurcation due to the HCA/SEA phase as well as highly-bound orbits.

The current BHBs in HMXBs, if capable of surviving the formation of the second BH (which in some of these HMXB systems there is a large chance of producing a HN/GRB event), have a reasonable chance of becoming a BBH which will merge during a future GW transient event (Apprndix H).

CE/GE evolution is an extremely-important and outcome-defining phase of binary- and multiple-stellar systems evolution which is still poorly understood.  
How these phases come to an end, in general, is one of its unsloved mechanisms.
Here I have shown mechanisms which solve the merging problem for systems involving COs which are capable of HCA/SAE at nearly BHL accretion rates during MT phases.  Thus, the implications for binary-evolution and population-synthesis numerical codes, and the results they produce are wide and important.  Predictions on rates for HNe/GRBs, short GRB/Kilonovae, ULXs, HLXs, as well as GW mergers, rely on these methods and thus, are affected by the results I have presented here.


\section*{Acknowledgements}

Useful discussions with F. De Colle, D. L\'opez C\'amara, A. Vigna-G\'omez, and T. Maccarone have benefited this work.
The author has made substantial use of arXiv and ADS services where all data and materials are available as indicated by the references in this work.

\section*{Data availability}

The data underlying this article will be shared on reasonable request to the corresponding author.

\section*{Appendix A:  On HN/GRB central engines}

Producing KBHs with natal spins is necessary to explain HN/GRB central engines, the leading mechanisms to do this rely on resupplying angular momentum through a stellar-binary channel (e.g., tides); in such cases, it is neceassary for the pre-BH binary to have a substantially less-massive (as compared to the BH progenitor star) companion so that the orbital separation may be small enough to spin the pre-collapsing star up to large AM values. 
\cite{2011MMB,2007Brown,2008Brown,2014MM} estimate the natal $a_\star$ of the BHs and energies that should have been available for a number of XBs in or near the Milky Way at the time of CC.  The IMXB LMC X-3 \citep{2008Brown}, with a modest-mass companion star, $M_{\rm comp} \sim 5 \msun$, was the likeliest GRB/HN progenitor, as its estimated natal spin of $a_{\rm \star,LMC X-3} \sim 0.45$, sets its energy budget at around $60 B$ (leaving behind the observed BH with spin of $a_\star \sim 0.25$), enough for a GRB/HN event, but unlikely to rapidly destroy the central engine before the energy can be used (Goldilocks scenario \cite{2008Brown}).  
In escence, given that the secondaries do no fit within short-enough orbits to spin the BH progenitors up, HMXBs should have much larger BH masses but much smaller BH spins ($a_\star \lesssim 0.05$, perhaps up to 0.2) and rotational energies.  So, not all BHs have to be born as SBHs, HN/GRB require some degree of KBH, but more likely from IMXBs, not for HMXBs.  
Of course, the descendant BBHs of IMXBs may be less likely to be detectable in a merger than those of HMXBs given the lower masses involved (and given the likelyhood of hosting a HN/GRB and the lower mass of their components, they are more likely to be disrupted at BH formation; however, they may well be more common and fit into more bound orbits, which could help compensate).

It is interesting to note that all three WR-BH HMXBs (IC 10 X-1, NGC 300 X-1 and Cyg X-3; Table 2) are already within the orbital separation (or obital period below two days) where the tidal-synchronization timescale is short (thousands of years or less) and can impart a substantial amount of angular momentum onto the WR star  before CC (see table 1).  Thus, it should be expected that these three systems produce a HN/GRB type of explosion (probably producing very eccentric orbits as opposed to breaking up given that the orbits are well bound and the WR are stripped, thus mass loss might not be significant), and, likely (accounting for masses and orbital binding energies), a future merger of two KBHs (if the BBHs survive) observable by a GW observatory.

\section*{Appendix B:  On the fallback model to explain the BH spins.}

The mechanism of spinning up BHs by SN fallback acretion of \citet{2017Batta,2018Schroder} seems capable of explaining BH spins in HMXBs of $a_\star \lesssim 0.8$.  It may still have problems with those around $a_\star \sim 1$.  
Some issues with this model explaining the HMXBs we are dealing with in this article are the following:;
\begin{itemize}
\item \cite{2019Fuller} point out that if the exploding star was H-stripped, there is little or no fallback.  These HMXBs have, most likely, gone through a CE phase where the primary star transferred most of its H-envelope onto the companion, thus the collapsing stars were in such a situation. However, the argument of \cite{2019Fuller} does not hold, as the mechanism is independent of the composition; i.e., the relevant lever arm for the torque is not from the H-envelope extension, but from the orbital separation, and if the He layer can be lifted to a radius beyond the orbital separation (the amount of material lifted may be relevant to the amount of specific AM required to be transported between orbit and spin, though), then that should be enough for the model in \cite{2017Batta,2018Schroder} to have merit.
\item More relevant may be the issue that the lifted-envelope material, having low velocity can be rapidly accreted by the companion (similar to wRLOF \cite{2007Mohamed,2018ElMellah,2019ElMellah}, where most of the material is capured by the companion). If the orbital velocity is similar or larger than the velocity of the lifted envelope it will be mostly swept and accreted by the larger-cross-section and more-massive companion, thus, effectively preventing accretion of mass and angular momentum by the BH.  
\item Perhaps the most important problem with the \cite{2017Batta,2018Schroder} model however, is that the prediction of the most relevant parameters are inverted with respect to HMXB distribution of these: From fig.3 in \cite{2017Batta}, as well as columns (5) and (6) in Table 1 of this work we can see that the relationship between BH spins to orbital separations is inverted; larger spins should be observed for shorter orbital periods.  This is not to say that, perhaps, a different set of parameters or configurations could resolve the issue or not, but that the most immediate predictions of the model do not fit the data; hence, at the very least, fine tunning seems to be necessary from the \cite{2017Batta,2018Schroder} model to fix this issue. 
\item In the simulations presented, none explain the largest BH spins in the observed HMXBs.  
\end{itemize}

\section*{Appendix C:  Implications for and from ULXs.}

If the HMXB systems in tables 2 \& 3 formed their BH about $10^6$ years ago, underwent a quick CE stopped by ablation or accretion as we propose, and continued transfering mass through RLOF  but mediated by a KH timescale (i.e., an intermitent or pulsed GE phase) at a rate that allows the current systems to transfer and accrete, i.e., some $10 \msun$ per Myr period, that gives an accretion rate of about $\dot{M} \sim 10^{-5} \msun$ yr $\sim 10^{21}$ g s$^{-1}$, which produces an average luminosity of $L \simeq 10^{42}$ erg s$^{-1}$ for the last Myr.   This is right where ultraluminous X-ray sources (ULXs) luminosities lie.  Such an accretion rate is borderline between the HCA and SEA regimes (within $10^3$ and $10^4$ Eddington).
In a sense, ULXs may be our view of these HCA GE HMXB systems as they transform their SBH into KBHs readying them for a future GW merger with large BHs spins.
As the HCA/SEA-fed BHs grow larger masses and spins, their jets luminosity must increase as well; thus, as these systems evolve and keep accreting from their companion stars, they may likely become brighter ULX sources with time (unless the excess goes mostly through the neutrino channel).  
Bubbles/cocoons detected in these systems could help put constraints on photon vs neutrino fluxes produced by the HCA/SEA on these systems, as well as if a continuous ULX stage of GE occurs or a, KH-timescale, intermitent ULX through pulsed GE dominates.

The HMXBs we include in this work, actually, show luminosities closer to $10^{38}$ erg s$^{-1}$ (Table 2, (10)). 
Most likely (and from direct observation of the current status of these HMXBs), accretion is wind fed most of the time, and, every $\tau_{\rm KH} \gtrsim 10^4$ years, an episode of mass transfer may occur, giving rise to a limited time period of HCA feeding frenezy and, perhaps, intermitent ULX activity.

\section*{Appendix D:  On Taylor-Spruit dial downs.} 

The Taylor-Spruit (TS) dynamo \citep{2002Spruit} is an important mechanism which regulates the angular transport inside stars. 
As such, it has been calibrated in stellar-modeling codes by using the observation of NSs, WDs and stellar (Sun) spins \citep[][respectively]{2005Heger,2008Suijs,2005Eggenberger}.  \citet{2019Fuller} discuss that the TS dynamo has in general been underused in stellar models if the population of low mass stars is to be explained.  Using this calibration, but for massive stars, they find that most single (or non-torqued up, e.g., by tides or mass transfer) BHs should be born with low spins.  However, by taking the Eddington-limit narrative, they also conclude that BHs in HMXBs are born with large spins (without providing a mechanism to acheive this).

\cite{2019Qin} assume in their models that, contrary to the results in \cite{2019Fuller,2005Heger,2008Suijs,2005Eggenberger}  (note that  \citet{2019Fuller} was released after \citet{2019Qin}), the TS dynamo can be dialed down in their massive-stellar models (winds are also set low compared to the typical values for the metallicities of the systems they describe, however, clumping can justify this assumption) in order to try to explain the observed large spins in the HMXB BHs; they produce nearly 5,000 models (including some with CHRM models), but do not fully reproduce the obesrvations (missing spins, orbital periods, etc.); they do not take into account the energy available in the BH progenitor.
In \cite{2014MM} a similar assumption was taken (dialing down TS and/or the poloidal magnetic field which connects the evolved stellar core with the stellar envelope) in the context stellar models for the progenitors of HN/GRB \cite{2011MMB,2007Brown,2008Brown}; nonetheless, this is required only for late and short stages of stellar evolution, after Case C mass transfer; something which is explicitly allowed by, and hence consistent with, the results in \cite{2019Fuller}.

\section*{Appendix E: On obtaining the results presented in Table 3.}

Conclusions reached by papers \cite{2019Qin} (row 1), \cite{2016MM} (row 2), \cite{2017Batta,2018Schroder} (row 3), and \cite{2008MM,2011MM} (row 4), and if missing, extrapolated from analytical energy estimates. Not all these papers are aimed at reproducing these HMXBs but only to explain their large spins (in fact \cite{2016MM} aims to show that natal kicks, through a directed SASI, cannot produce such large spins), so they are not expected to match the observations.
Estimates derived from the conclusions of this paper, and applied to the conclusions of \cite{2019Qin} are presnted in row 5; the conclusions of this paper are applied to the model in \cite{2008MM,2011MM} to acheive row 6.

\section*{Appendix F: Evolutionary paths and implications for LIGO sources.}

Systems like M33 X-7, Cyg X-1, LMC X-1, IC 10 X-1, and likely NGC 300 X-1 and Cyg X-3, are likely formed with a massive primary which transfers a substantial amount of its envelope mass to their less massive companion star (i.e., up to a few tens of $\msun$) and rejuvenates it (thus, increasing its lifespan by up to a few Myrs \cite{1995Braun}) in either, late-Case A (late during MS \footnote{Let us remember that a very massive Fe core must be allowed to form before CC; so this scenario is not highly favored, especially in the larger metallicity systems, as winds from a stripped stellar He core will probably allow the star to lose too much mass and produce a neutron star as opposed to the massive BHs observed; however, this could explain the apparent BH-mass diference in Cyg X-3 (Table 2) with respect to the other HMXBs.}), case B (during the RG phase; once most of the He mass of the star has been produced, $\sim90\%$, and the formation of a BH is more likely; also, close to 90\% of the radial expansion occurs at this stage, so this has the largest probability considering the possible initial orbital separations between these binaries), or even better, Case C (here the He core has grown even larger), favoring the BH-mass observations (Figure 1 shows a general schematic outline of this stellar evolutionary set of channels for the formation of HMXBs until they reach said stage; Table 2, columns (2,3) show the masses in the HMXBs).  
The latter mechanism has the two MT runaway processes active, so it may also be the one that better explains the final, small orbital separations.
This is the case discussed in \cite{2008MM,2011MM,2007Brown,2008Brown,2011MMB,2014MM} given that it allows LMXB and IMXB to obtain and retain the largest amount of angular momentum for HN/GRB central engines (Appendix A).  
Prior to the mass transfer, $q > 1$ (the primary star, or progenitor of the current BH in said systems, being more massive); then, the nearly conservative MT via RLOF and CE strips the H envelope from the primary star and onto the secondary, inverting $q$ (helping bring this first CE phase to an end), decreasing the original orbital separation which can be as large as several-hundred to over-a-thousand $\rsun$, to only a few $\rsun$, an orbit not much wider than those currently observed in these HMXBs (particularly, the O-star systems).  
The two stars in these systems, being rather massive and not-too-compact, do not fit in orbits with periods shorter than several days, thus, even if they tidally synchronize, not enough angular momentum can be transferred from their orbit to their spins \cite{2008MM,2011MM}.
Thus, a massive-stellar SBH forms in a rather dark CCSN or by direct collapse.  
The system is hardly affected due to the mass loss during the SN (from Blaauw-Boersma -BB- kicks, \cite{1961Blaauw,1961Boersma}, we know the system cannot lose more than half the mass before the SN, otherwise the system is unbound) mostly because most of the envelope has been transferred onto the companion star which helps keep the binary bound (along with the substantial decrease in orbital separation).
The fact that the core is very massive helps by reducing the likeliness of the SN being successful, thus favouring a direct collapse into a massive stellar SBH. 

After CC of the primary into a massive-stellar SBH, the secondary star will wRLOF, and eventually RLOF staring a CE/GE/pulsed-GE phase (the one where we have descibed, in section~\ref{sec:CEsep}, how to avoid or stop the merger; the case of the pulsed-GE will alternate with extended KH-periods of wRLOF and it is likely this is the current situation in most of these HMXBs) of MT back onto the BH which spins it up to a KBH (to the presently observed values) by increasing the BH mass by a factor of 1.5 to 2.5\footnote{This depends on the final spin, as well as the location of the ISCO: at the marginally-stable or marginally-bound orbit; Fig. 6 \cite{2000Brown}.}, and which may further decrease the orbital separation (as $q > 1$, again, although, note that here the originally secondary star has become the massive donor component of the system).
If this is the case, and if the orbits shrink to orbital periods within 1 to 2 days (or less), the secondary stars will probably synchronize with the orbital period within a few kyrs \citep{1977Zahn,1975Zahn,1989ZahnA,1989ZahnB,2011MMB,2007Brown,2008Brown,2014MM}.  
In the case of the WR HMXBs in this study, it is likely that these systems underwent a CE phase towards a merger, which was accretion-separated, thus MT was nearly conservative, allowing for a further decrease of the orbital separation (the radius of the BH no longer being an issue) up to the point where $q \sim 1$ and are either synchrinizing or fully synchronized.
If this is the case, they will most likely produce a central engine and, thus, explode  as HN/GRB explosions when the second BH forms.  
However, if the mass transferred onto the massive KBHs is large, or if little mass is lost during the HN (again, the three WRs are much less massive than the O stars in the other three HMXBs), the BB kicks will not unbind these systems (the HN kicks could still do this if not properly aligned, though; but the orbits are more bound); the tighter orbital separation may also help keep the binary from breaking appart (but the eccentricity may become extreme).  
In any case, if these binaries survive the formation of the secondary BHs (regardless of whether or not they produce HN/GRB explosions) and if the SNe and BB kicks are favorable, they have the possibility, if the final orbital separatations are within a couple of tens of $\rsun$, to form future BBH mergers detectable as gravitational waves (GWs) events.

Cyg X-1 has a spin which is extremely close to $a_\star \simeq 1$ and the cosmic censorship hypothesis may be at play, it also has an orbital period which is currently the largest among the six HMXBs, so it is questionable whether the BH may easily accept any more mass (unless forced into a CE) and thus, whether the orbital separation may grow any smaller (e.g., driven into a merger attempt which separates by CE-separation by accretion).
Hence, it is unlikely that a central engine may form.  
Thus, a massive SBH may form when the secondary collapses.  
Due to the large mass of the companion star in Cyg X-1, it may produce another BH around  $M_{BH} \sim 20 \msun$, not unlike the BBH systems in GWs $190408\_181802$, $190512\_180714$, $190708\_232457$, $190719\_215514$, or $190828\_065509$ \cite{2021Abbott}.
However, its current orbital situation does not favor it within a Hubble timescale (HT); nonetheless, a correctly aimed SN kick may help. 

Given the current mass and apparent age of the companion in M33 X-7, the BH may accrete a further 10 to 15 $\msun$ from its secondary star spinning it up to $a_\star \simeq 1$, and this may still allow for the secondary to collapse into a massive 30-$\msun$ BH. 
If so, it could lead to the formation of a system analogue to that which gave rise to GW150914.  
Given the size, mass, $q$, and MT necessary to spin the BH, it is unlikely that the secondary BH will form as a KBH (unless a merger attempt with CE-separation by accretion phase can take place).  
Thus, the binary has big chances of surviving such an event. 
Now, unless the orbital period can manage to get shorter than a couple days (by MT or SN/BB kicks) it may not merge by GW emission within a HT. 

The BH in LMC X-1 could very well double its mass before it reaches its maximum spin and brings $q \lesssim 1$. Its orbit like that of Cyg-1 and M33 X-7 does not currently help predict a high-likelyhood merger within a HT; but once more, SN kicks or a CE-separation by accretion scenario could help produce such an outcome.

The cases of IC 10 X-1, and probably NGC 300 X-1, dependending on their actual masses, may be somewhat similar (as their companion types and orbital periods are), they may form HN/GRB events, but their strongly bound orbits may help them survive, and in the case of low mass loss, they are likelier candidates for BBHs that may merge by GW emission within a HT.

The outlayer among the systems we study is Cyg X-3.  It may have the smallest masses, but these still remain uncertain. What is clearly different is that the system is extremely bound at $P_{orb} \sim 0.2$ days, allowing little doubt of tidal synchronization and, most likely, a HN/GRB local (Galactic) candidate within the next Myrs (even if the CC mass were low, it could esaily produce a millisecond magnetar as a central engine).  Its tight orbit will most likely allow it to also produce a GW event in the not-so-distant future, unless the HN/GRB event is powerful enough to break it appart even in such a tight binary; formation of a BH rather than a magnetar might also help keep the system bound.

An interesting insight here is that, given the KBHs already present in the O-star systems, we can infer such KBHs present as well in the WR systems, and that the latter WR stars are probably synchronizing/synchronized.  Central engines will develop during the CC of the secondaries of at least the WR systems, but being tight binaries having lost the H envelopes, and not being too massive, they may turn into BBH with two KBHs which may merge within a Hubble timescale.  However, the BH spins, and orbital AM vector, may be arbitrarily oriented, even if they are originally parallel (as may be expected due to the formation channel here suggested).

If MT can proceed close to conservative while still preventing a merger (through the CE accretion-separation mechanism) in the three O-star-BH HMXBs (M33 X-7, Cyg X-1, and LMC X-1) and the orbital separations decrease, they may evolve into WR-star HMXBs similar to the other three systems (at larger masses and orbital separations, albeit).  If these are indeed two stages of a similar evolutionary track, this is another reason to suspect that the expected Kerr parameters of the current BHs in NGC 300 X-1 and Cyg X-3 may be as high as those in the rest of the HMXBs we have discussed.

\citet{2021Klencki,1997Soberman,1994Shore} argue that in order to remove a CE phase, it is much more favorable to have a convective envelope (like in Case C mass transfer) as it expands at the loss of material, so it is less bound.  If the envelope is radiative, it contracts, and becomes more bound and it is harder to remove, possibly leading to a merger in the case of HMXBs under the $\alpha\lambda$ CE prescription, which would otherwise evolve into the BBHs which merge by GW emission and are detectable by LIGO.
However, in our BHL HCA regime the BH may fully accrete an envelope if it is not unbound, i.e., if it is radiative, (during late MS or RG, case A or B mass transfer), and instead, case C will remove a larger portion of them (which could result in lower accretion into the BH).
Assuming a merger is regularly prevented, a bifurcation in masses and spins could occur depending on whether Case A or B vs Case C mass transfer are the cause of a CE/GE phase.
By preventing mergers the prescriptions presented in this article turn around the conclusions of the $\alpha\lambda$ CE prescription  that BBHs cannot be formed by CE in a radiative envelope (Cases A or B), and replaces it with the possible bifurcation in BH mass distribution. 
The BH masses and BH spins in the predictions of many binary evolution channels for BHBs (one BH) and BBHs (two BHs) might need to be revisited and modified accordingly as well.

Out of the binary stellar evolutionary channels to produce BBH LIGO sources, HMXBs are among the main contenders.  LMXBs will not form a second BH by definition.  IMXBs, may produce CO-WD, CO-WD (CO stands for compact object; these may not be LIGO sources but are GW sources nonetheless), or BH-NS sources which have been observed in a few events (see, e.g., GW200105 and GW200115\footnote{Where no electromagnetic counterparts were detected, see e.g., \cite{2021Becerra}, which is, by itself, an interesting result given the masses, $q$, and, hence NS and BH radii involved.} \cite{2021Abbott}), however, having the main HN/GRB sweetspot in the BH- spin/energy range, many binaries might get disrupted instead.  HMXBs, which currently are BHBs, can naturally evolve into smaller-orbit BH-NSs or BBHs which may merge in a Hubble timescale.  The ones with the most bound orbits during the BHB-stage could potentially develop HN/GRB central engines which will produce a bright tranient source and could break appart (but given the masses involved are less prone than those in IMXBs), the rest of them may be well poised to explain a large portion of the LIGO sources (along with the channels coming from multiple systems and/or dynamical exchanges in clusters).

\section*{Appendix G: Possible observational tests.}

A possible observational confirmation of a KH-mediated HCA/SAE regime at work in HMXB or ULX systems would consist on identifying these by observing their nebula in search for {\it clumps or nodules}, if their velocity is fast with respect to the ambient medium, or bubbles within bubbles if they are static.  Estimating the energy required to inflate each bubble/nodule, and their ages (from their velocities) can help differntiate HCA/SAE stages from wind feeding stages.  
If accretion rates are close to 10$\%$ of BHL ($\sim 1 \msun$ day$^{-1}$), as expected from the dynamical, unstable MT due to their $q$, and jets form to separate the CE by ablation with an efficiency of 0.1\% (acounting for accretion efficiency of 10\% plus photon to neutrino efficiency of 1\%) we obtain $L_{jet} \sim 10^{46}$ erg s$^{-1}$ over $10^5$ s, thus lifting a $10^{51}$ erg bound CE in about a day.  KH timescale is of order $10^{4-5}$ years for the massive companions in the HMXBs in Tables 2 and 3, so, in a Myr timescale there should be 10 to a hundred day-long outbursts (depending on actual sizes of the MT.

ULXs are interesting since, at $10^{39} \lesssim L_x \sim \lesssim 10^{42}$ erg s$^{-1}$ they cover the SEA range for a $10-\msun$ BH.  So orbital modulation of days and orbital velocities of $v_{orb} \sim 10^7$ cm sec$^{-1}$ could confirm the existence of HMXBs within
the accretion range we explore in this article.
Obesrvartions of sources with larger luminosities, e.g., up to 4 or 5 orders of magnitude larger hyperluminous X-ray sources (HLXs) but similar orbital modulation and/or orbital velocities, extend that range well into the HCA range. \cite{2011Lasota} have observed variability in HLX-1 (at $P \sim 10^7$ s) and proposed a model for an Eddington-limited BH of $M_{BH} \sim 10^4$ which is fed by a star in an eccentric ($e \sim 0.3$) $P_{orb} \sim 380$ day orbital period.  We note here that a Cyg X-1-like or heavier system, undergoing a HCA/SEA GE during Case B or C (the latter would probably provide for larger accretion rates), with orbital separation of 6 to 7 AU may fit the dynamical model as well.  Thus HLX-1 could be the transition of a massive stellar SBH into a KBH in a HMXB. 
Further study would be interesting in this direction.

Systems where CE are observed may also help, but in such cases, the CE itself may interfere with obtaining $L_x$, nonetheless, if orbital parameters, velocities, and energies can be extracted, progress can be made as well.

\section*{Appendix H:  If the KBHs were natal.} 

By disregarding again, and for the last time, only for the sake of argument, the whole (fundamental and most important part of this whole excercise) energy argument against the natal large spins of the BHs in these HMXBs, we may still ask, why would such exotic mechanisms as I present in this article (HCA, CE, and Jets and/or full CE accretion) be required to explain the formation of such systems if something like the Case A mass transfer, or even CHRM systems explained in, e.g., \cite{2019Qin} {\it seem} to work?  
Well, because they do not really solve the problem of forming such large natal BH spins satifactorily either.  
They also need assumptions which do not match well with the observations.
It is further, highly unlikely that, inside such a kinetically-energetic star, differential rotation, convection, and thus, dynamos,  will not actively use the TS mechanism to transfer angular momentum within it during their MS.  
This situation must be maintained with this rotation profile for a few million years.  
The scenarios in their parameter space run into trouble even considering chemically-homogeneous, rotationally-mixed (CHRM) stars as they do not reach the observed orbital parameters either (some of the problems for CHRMs to produce such BH spins in HMXBs were studied and pointed out as early as \cite{2008MM} and later in \cite{2011MM}).

To finish this hypothetical arc let us finally return to answer one of the questions we left pending:  If these BHs were born with such large spins and energies, what would the expected observable consequences be for CC?
The precollapsing stars in these HMXBs systems had their $10^{53}$ energy budget from a typical CCSN, but, if the observed BH spins where natal, they had between $10^{54-55}$ ergs of extra {\it kinetic} energy available during collapse (so, it would not be as easily lost to neutrinos).  
During CC, a large fraction of the material will not be able to fall directly onto the PNS or BH in the center, but it will rather form a massive accretion disk \citep[as discussed in][as well as other literature on collapsars]{1999MacFadyen,2011MMB,2007Brown,2008Brown,2014MM} and a large magnetic field; a Penrose process \cite{1971Penrose} will most likely emerge, e.g., a BZ (Blandford and Znajek, \cite{1977BZ}) and/or a BP (Blandford-Payne, \cite{1982BP}) central engine, producing a GRB and a HN, blowing the rest of the star appart and, likely, preventing the formation of a very massive stellar BH (so, it is to be expected that $M_{BH} < 10 \msun$) .  
This is not consistent with what observations show, all BHs in these HMXBs are rather heavy (with the possible exception of CygX-3, all have $M_{BH} > 10 \msun$, see (3) in Table 1); Cyg X-1 had a dark explosion or no explosion at all \cite{2003Mirabel}.  
None of the companion stars show strong HN pollution, although, this is difficult to corroborate.  The eccentricity seems to be small in all these systems (eventhough the orbital separations are large so tidal circularization is of order Myrs \cite{1977Zahn,1975Zahn,1989ZahnA,1989ZahnB}).  
No strong evidence of large spin (or jets) missalignment with the orbital-angular-momentum vector. 
Thus, it is highly unlikely that the evolutionary paths which assume a large natal spin are correct.  
This is not to say that all of these systems came out of direct collapse onto BHs nor from dark explosions, only that, at most, the SNe explosions were modest at best, the natal spins were originally much smaller than they currently are ($a_\star < 0.2$ as pointed out in \cite{2016MM,2008MM,2011MM}) and it is unlikely that a central engine formed in any of these HMXB systems during the collapse that formed their, originally SBHs, currently KBHs.  
It is important to note then, that if a CE/GE phase has occurred, with the implied mass and angular-momentum transfer, this naturally explains the spin-orbit alignment (even if it was not the case after BH formation), the low eccentricity, the low peculiar velocity (if SN is small or abscent), and perhaps the low pollution of companion (with caveats).  



\end{document}